\documentstyle[12pt]{article}

\input{tcilatex}
\begin{document}

\title{{\bf Noncommutative Geometry Framework and The Feynman's Proof of Maxwell
Equations}}
\author{A. Boulahoual and M.B.Sedra\thanks{%
Corresponding author: sedra@ictp.trieste.it} \\
International Centre for Theoretical Physics, Trieste, Italy}
\maketitle

\begin{abstract}
The main focus of the present work is to study the Feynman's proof of the
Maxwell equations using the NC geometry framework. To accomplish this task,
we consider two kinds of noncommutativity formulations going along the same
lines as Feynman's approach. This allows us to go beyond the standard case
and discover non-trivial results. In fact, while the first formulation gives
rise to the static Maxwell equations, the second formulation is based on the
following assumption $m[x_{j},\dot{x_{k}}]=i\hbar \delta _{jk}+im\theta
_{jk}f.$ The results extracted from the second formulation are more
significant since they are associated to a non trivial $\theta $-extension
of the Bianchi-set of Maxwell equations. We find $div_{\theta }B=\eta
_{\theta }$ and $\frac{\partial B_{s}}{\partial t}+\epsilon _{kjs}\frac{%
\partial E_{j}}{\partial x_{k}}=A_{1}\frac{d^{2}f}{dt^{2}}+A_{2}\frac{df}{dt}%
+A_{3},$ where $\eta _{\theta }$, $A_{1}$, $A_{2}$ and $A_{3}$ are local
functions depending on the NC $\theta $-parameter. The novelty of this proof
in the NC space is revealed notably at the level of the corrections brought
to the previous Maxwell equations. These corrections correspond essentially
to the possibility of existence of magnetic charges sources that we can
associate to the magnetic monopole since $div_{\theta }B=\eta _{\theta }$ is
not vanishing in general.
\end{abstract}

\hoffset=-1cm \textwidth=11,5cm \vspace*{1cm}

\hoffset=-1cm\textwidth=11,5cm \vspace*{0.5cm}

\newpage

\section{Introduction}

Noncommutative geometry (NCG) stimulated recently by Connes [1] and
developed later on by several pioneering authors [2, 3] have played an
increasingly important role more notably in the attempts to understand the
space-time structure at very small distance. Much attention has been paid
also to quantum field theories on NC spaces, in particular NC Yang-Mills
gauge theory as well as NC-QED, a subject which has matured as an area of
intense research activity in more recent times [4-6]. In fact it has been
established by Seiberg and Witten [2] that the existence of noncommutativity
in open string boundaries in the presence of the NS-NS B field results in NC
D-branes to which the open string endpoints are attached. Related to these
stimulating ideas, a wide number of papers were devoted to study several
aspects of the NC [7 ]. \newline
One particular property of NCG framework is its richness and also the fact
that we can discover the standard results just by requiring the vanishing of
the deformed parameter which means also the vanishing of noncommutativity.
Note that the passage from commutative to NC space time is simply achieved
by replacing the ordinary product, in the space of smooth functions on $R^{2}
$ with coordinates ($x,t),$ by the NC associative $*$ product. Works having
used this NC formalism are various and the results found are numerous, we
will limit ourselves to mention some of them, namely [8-14].\newline

The aim of this paper is to study another aspect of the noncommutativity
framework adapted to the Feynman's proof of Maxwell equations [15-19]. As
well known, a century ago, Maxwell brought four basic laws dealing with
electromagnetism, these laws describe the evolution in time and space of the
electric and magnetic fields $E$ and $B$. Together with the Lorentz force
law, the Maxwell equations constitute a complete description of
electromagnetism (the Maxwell theory). Furthermore these equations have
different forms, vectorial, differential and can be proved in different way.
Feynman in 1948 has given a proof of these equations, assuming only Newton's
law of motion and the commutation relation between position and velocity for
a single nonrelativistic particle. The importance of this proof emerged
notably with the Dyson's work [15]. As signaled in this work, the motivation
of Feynman was to build a new theory outside the framework of conventional
physics, but his assumptions using these commutation relations and the
Newton's equation were not lead to new physics [15]. This proof, although
based on simple mathematical assumptions, is shown to give rise to
nontrivial generalizations [16-19].\newline

Among many possible existing extensions, we are going to adapt thereafter
the NC framework to the Feynman's proof, a fact which leads us to extract
important results. We present two kind of NC formulations and show in a
first one that the application of the Feynman's proof in NC space, leads to
the static Maxwell equations. Focusing to obtain a new theory, we propose in
our second formulation to modify the Moyal bracket between the position $%
x_{i}$ and the velocity $\dot{x_{j}}$. This task can be easily accomplished
by assuming that the velocity is space dependent and then the star product
between $x_{i}$ and $\dot{x_{j}}$ becomes non trivial. This assumption will
modify the Maxwell equations, giving rise to a new theory where extra terms
proportional to NC $\theta $ -parameter appear. The results extracted from
the second formulation are more significant as they are associated to a non
trivial $\theta $-extension of the Bianchi-set of Maxwell equations namely $%
div_{\theta }B=\eta _{\theta }$ and $\frac{\partial B_{s}}{\partial t}%
+\epsilon _{kjs}\frac{\partial E_{j}}{\partial x_{k}}=A_{1}\frac{d^{2}f}{%
dt^{2}}+A_{2}\frac{df}{dt}+A_{3},$ where $\eta _{\theta }$, $A_{1}$, $A_{2}$
and $A_{3}$ are local functions depending on the NC $\theta $-parameter.%
\newline

Our objectives in reconsidering the Feynman's proof are, on one hand, to put
it in relief and, on the other hand, to show its importance in the NC
framework. The novelty of this proof formulated in the NC space is revealed
notably at the level of the corrections brought to the standard Maxwell
equations. These corrections correspond essentially to the possibility of
existence of sources of magnetic charges that we can associate to the
magnetic monopole since $div_{\theta }B=\eta _{\theta }$. Note that these
extra terms $\eta _{\theta }$ are absent in the ordinary case associated to $%
\theta =0.$ These results may give new insights into the study of the
electromagnetic duality and its various physical and mathematical aspects.%
\newline

This paper is organized as follows. In section 2, we summarize some
properties of the Poisson manifold and review the Feynman's proof of the
Maxwell equations. In section 3 we present some useful identities of $*$
product after that we examine the Feynman'proof of Maxwell equations in NC
spaces. Section 4 is devoted to our concluding remarks.

\section{Maxwell Equations: The Feynman's Proof}

Maxwell equations have played a pioneering role in physics and they continue
to nourish several axes of research either in physics or in mathematics.
Their formulations as well as the survey of their solutions constitute a
topic of a big interest [20] and it's in this context that are located the
famous theories of Yang-Mills. Recently, we attended to a new approach
leading to the derivation of these equations and based on what is called the
Feynman's proof of Maxwell equations. Details concerning this approach are
presented in the Dyson's work [15]. Later on, several authors took this
approach and tried to put in relief the Feynman's idea and to develop it or
sometimes to generalize it to other contexts [16-19]. Before reviewing the
Feynman's proof of the Maxwell equations, lets first start by introducing
some basic algebraic properties of the underlying Poisson manifold ${\cal M}$

\subsection{Some algebraic properties}

In fact the previous approach can be simply stated in the general way as
finding all Poisson tensors on a phase space manifold such that they have
Hamiltonian vector fields which correspond to second order differential
equations such that $\{q^{i},q^{j}\}=0,$ with the symbol $\{,\}$ standing
for the Poisson bracket defined usually as

\begin{equation}
\{f,g\}=\frac{\partial f}{\partial q^{i}}\frac{\partial g}{\partial p_{i}}-%
\frac{\partial f}{\partial p^{i}}\frac{\partial g}{\partial q_{^{i}}}
\end{equation}

where $f$ and $g$ are two functional of $q$ and $p$. Denoting by ${\cal A}$
the algebra of classical observables on the manifold ${\cal M},$ one can
define a Poisson structure $\{,\}:{\cal A\times A}\rightarrow {\cal A}$ on
this manifold as been a skew-symmetric bilinear map such that:

a) $({\cal A}$ , $\{,\})$ satisfies the Jacobi identity\smallskip
\begin{equation}
\{F,\{G,H\}\}+\{H,\{F,G\}\}+\{G,\{H,F\}\}=0
\end{equation}
b) The map $X_{F}=\{.,F\}$ defines a derivation on ${\cal M}$ of the
associative algebra ${\cal A(M)}$, that is, it satisfies the Leibnitz rule
\begin{equation}
\{F,GH\}=G\{F,H\}+\{F,G\}H
\end{equation}
A manifold endowed with such a Poisson bracket on ${\cal A(M)}$ is called a
Poisson manifold. Furthermore, consider a Poisson manifold ${\cal P}$, for
any $H\in {\cal A(P)}$, there is a unique vector field $X_{H}$ on ${\cal P}$
such that
\begin{equation}
X_{H}G=\{G,H\}
\end{equation}
for all $G\in {\cal A(P)}$. $X_{H}$ is nothing but the Hamiltonian vector
field of $H$. Now one can define a dynamical system on the Poisson manifold $%
{\cal M}$ just by considering for any function $H\in {\cal A}$ the following
differential equation
\begin{equation}
\frac{dF}{dt}=\{F,H\}
\end{equation}
Moreover, one can also express the Poisson bracket $\{F,G\}$ in any set of
local coordinates $(x^{a})$ in the following way
\begin{equation}
\{F,G\}=X_{G}F=\{x^{a},G\}\frac{\partial F}{\partial x^{a}}
\end{equation}

\subsection{The Feynman's Proof.}

This section is devoted to an explicit remind of the main steps involved
into the Feynman's proof of the Maxwell equations in their classical form
[15, 16]. Our objective is to present these calculations in order to make a
comparison with the NC case to be discussed later. This proof is essentially
based on the Newton's laws of the non relativistic classical mechanics and
on the relations of commutation joining the coordinates of position and
velocity of a single non relativistic particle. An extension to the
relativistic case is naturally possible [17, 19] and may leads to important
results more notably in connection with quantum field theory approaches. The
manifold we consider is parameterized by local coordinate variables $%
(w^{a})=(x^{i},\dot{x}^{i})$ of a non relativistic particle whose position $%
x_{j}(j=1,2,3)$ and velocity $\dot{x_{j}}$ satisfy the Newton's equation

\begin{equation}
m\ddot{x_j}=F_{j}(x,\dot{x},t) ,
\end{equation}
with commutation relations
\begin{equation}
\begin{array}{lcr}
\{x_{j},x_{k}\}=0, &  &
\end{array}
\end{equation}
\begin{equation}
\begin{array}{lcr}
m\{x_{j},\dot{x_{k}}\}=i\hbar \delta _{jk}. &  &
\end{array}
\end{equation}
Then, there exist a couple of fields $E(x,t)$ and $B(x,t)$ that we can
identify with the electric and the magnetic fields respectively such that we
get the Lorentz force law
\begin{equation}
F_{j}=E_{j}+\epsilon _{jkl}\dot{x_{k}}B_{l},
\end{equation}
and the first couple of the Maxwell equations
\begin{equation}
divB=0,
\end{equation}
\begin{equation}
\frac{\partial B}{\partial t}+\nabla \times E=0.
\end{equation}
The second couple of Maxwell equations
\begin{equation}
divE=4\pi \rho ,
\end{equation}
\begin{equation}
\frac{\partial E}{\partial t}-\nabla \times B=4\pi j,
\end{equation}
merely defines the external charge and the current densities $\rho $ and $j$
respectively.\newline
The Feynman's proof starts by differentiating the bracket (9) with respect
to time and using (7), we have
\begin{equation}
\{x_{j},F_{k}\}+m\{\dot{x_{j}},\dot{x_{k}}\}=0.
\end{equation}
Then using the Jacobi identity
\begin{equation}
\{x_{l},\{\dot{x_{j}},\dot{x_{k}}\}\}+\{\dot{x_{j}},\{\dot{x_{k}}%
,x_{l}\}\}+\{\dot{x_{k}},\{x_{l},\dot{x_{j}}\}\}=0
\end{equation}
as well as the bilinearity of the Poisson bracket we find the following
constraint equation
\begin{equation}
\{x_{l},\{x_{j},F_{k}\}\}=0.
\end{equation}
Furthermore, since the bracket is antisymmetric the tensor $\{x_{j},F_{k}\}$
satisfy
\begin{equation}
\{x_{j},F_{k}\}=-\{x_{k},F_{j}\},
\end{equation}
and therefore we may write
\begin{equation}
\{x_{j},F_{k}\}=-i\frac{\hbar }{m}\epsilon _{jkl}B_{l}.
\end{equation}
This equation gives a definition of the field $B$ whose components are $%
B_{l} $. This shows that $B$ would in general depend on coordinates $x,$ $%
\dot{x}$ of the Poisson manifold ${\cal M}$ and possibly time $t$. Combining
(17) with equation for $B_{l}$ (19) lead to
\begin{equation}
\{x_{l},B_{m}\}=0.
\end{equation}
On account of the basic equations (8-9), this means that $B$ is a function
of the coordinates $x$ and $t$ of the particle. Therefore, its shown that
the vectors $E$ and $B$ are not independent as we have
\begin{equation}
\{x_{m},E_{j}\}=0,
\end{equation}
which says that $E$ is also a function of $x$ and $t$ only.\newline
Now we have two equations (15) and (19) that we naturally need to compare.
The way to do this consist simply in writing $B$ as

\begin{equation}
B_{l}=-i\frac{m^{2}}{2{\hbar }}\epsilon _{jkl}\{\dot{x_{j}},\dot{x_{k}}\}.
\end{equation}
Another application of the Jacobi identity gives
\begin{equation}
\epsilon _{jkl}\{\dot{x_{l}},\{\dot{x_{j}},\dot{x_{k}}\}\}=0,
\end{equation}
leading naturally to the first Maxwell equation $divB=0$ (11) namely
\begin{equation}
\{\dot{x_{l}},B_{l}\}=0.
\end{equation}
Indeed,
\begin{eqnarray}
\{B_{l},\dot{x_{l}}\} &=&\{x_{a},\dot{x_{l}}\}\frac{\partial B_{l}}{\partial
x_{a}}  \nonumber \\
&=&\frac{i\hbar }{m}\frac{\partial B_{l}}{\partial x_{a}}\delta _{al} \\
&=&\frac{i\hbar }{m}divB  \nonumber \\
&=&0.  \nonumber
\end{eqnarray}
The proof of the second Maxwell equation (12) starts from deriving both
sides of (22) with respect to time. This gives
\begin{equation}
\frac{\partial B_{l}}{\partial t}+\dot{x_{m}}\frac{\partial B_{l}}{\partial
x_{m}}=-\frac{im^{2}}{\hbar }\epsilon _{jkl}\{\ddot{x_{j}},\dot{x_{k}}\}.
\end{equation}
Now by virtue of (7) and (10), the right side of (26) becomes
\begin{eqnarray}
-\frac{im}{\hbar }\epsilon _{jkl}\{E_{j}+\epsilon _{jab}\dot{x_{a}}B_{b},%
\dot{x_{k}}\} &=&-\frac{im}{\hbar }\left( \epsilon _{jkl}\{E_{j},\dot{x_{k}}%
\}+\{\dot{x_{k}}B_{l},\dot{x_{k}}\}-\{\dot{x_{l}}B_{k},\dot{x_{k}}\}\right)
\nonumber \\
&=&\epsilon _{jkl}\frac{\partial E_{j}}{\partial x_{k}}+\dot{x_{k}}\frac{%
\partial B_{l}}{\partial x_{k}}-\dot{x_{l}}\frac{\partial B_{k}}{\partial
x_{k}}+\frac{im}{\hbar }B_{k}\{\dot{x_{l}},\dot{x_{k}}\}.
\end{eqnarray}
On the right side of (27), the last term is zero by virtue of (22), the
third term vanishes also as it describes exactly the first Maxwell equation.
Now identifying l.h.s and r.h.s of (26), we get
\begin{equation}
\frac{\partial B_{l}}{\partial t}=\epsilon _{jkl}\frac{\partial E_{j}}{%
\partial x_{k}},
\end{equation}
which is nothing but the second Maxwell equation (12).\newline
This is the way followed by Feynman to prove the Maxwell equations in their
classical form. His motivation was to ``discover a new theory not to
reinvent the old one'', but the proof showed him that his assumptions (7-9)
were not leading to new physics. As was the case for several authors who
find interesting the Feynman's approach, we project in the forthcoming
section to go beyond this approach and setup the Feynman's proof in a
non-commutative space. The way to apply the noncommutativity is by replacing
ordinary product by star product and Poisson bracket or ordinary commutators
by the Moyal bracket.

\section{The Feynman's proof in the NC geometry framework}

\smallskip The passage to NC geometry, based essentially on the
noncommutativity of space-time coordinates, is justified among others by its
importance in different currents of research more particularly in high
energy physics. The deep idea behind the noncommutativity of coordinates is
that in a certain microscopic regime our standard conception of the
space-time is not more applicable. Such a regime is characterized by domains
of area $\theta $ where the space-time loses its condition of continuum and
becomes subject to the following new structure $[x_{\mu },x_{\nu
}]_{*}=x_{\mu }*x_{\nu }-x_{\nu }*x_{\mu }=i\theta _{\mu \nu },$where $%
\theta _{\mu \nu }$ is a real antisymmetric constant matrix and $[$.$,$.$%
]_{*}$ is the Moyal bracket. One way of incorporating noncommutativity of
coordinates in the context of field theory is through the Moyal product
based on the $*$-product that will be introduced latter on. To avoid hard
notations, we will later on simply denote the Moyal bracket by $[$.$,$.$].$
Before going on, let us first recall briefly some useful identities of the $%
* $-product.

\subsection{Some properties of $*$-product}

Recently the star product marks a remarkable success due to its intervention
in different aspects of string theory greatly related to NC geometry. In
this section, we give some useful properties of this star product as well as
of the Moyal bracket [9,10]. To define this object, lets start by
considering two functions $f(x)$ and $g(x)$ such that

\begin{equation}
f(x)*g(x)=e^{\frac{i}{2}\theta ^{ab}\frac{\partial }{\partial \xi ^{a}}\frac{%
\partial }{\partial \eta ^{b}}}f(x+\xi )g(x+\eta )/\xi =\eta =0,
\end{equation}
where $\theta ^{ab}$ is a constant, of dimension $[L]^{2}$, known as the NC
parameter\footnote{%
In all the parts of this paper the parameter $\theta $ is considered as a
constant matrix.}. This formula leads naturally to what is often called the
Moyal bracket of functions.
\begin{equation}
\lbrack f(x),g(x)]=f(x)*g(x)-g(x)*f(x).
\end{equation}
According to this definition, the commutation relation for the space
coordinates becomes
\begin{equation}
\lbrack x_{i},x_{j}]=i\theta _{ij}.
\end{equation}
Such a structure describes a NC space for which the space coordinates are
not necessarily commuting. Note by the way that the function $f(x)$ may
depend on space coordinates as it can depend on space-time coordinates.%
\newline
We collect here bellow some useful properties.\newline

1){\bf Associativity\newline
}
\begin{equation}
(f*g)*h=f*(g*h).
\end{equation}

2){\bf Jacobi identity\newline
}
\begin{equation}
\lbrack f,[g,h]]+[g,[h,f]]+[h,[f,g]]=0.
\end{equation}

3){\bf Leibnitz rule\newline
}
\begin{equation}
\lbrack f,g*h]=g*[f,h]+[f,g]*h.
\end{equation}

4){\bf Linearity\newline
}
\begin{equation}
f*(g+h)=(f*g)+(f*h).
\end{equation}
The star product is also compatible with integration
\begin{equation}
\int Tr(f*g)=\int Tr(g*f),
\end{equation}
where $Tr$ is the ordinary trace of the $N\times N$ matrices, and $\int $ is
the ordinary integration of functions.\newline
Another useful identity is given in term of local coordinates ($x^{a}$). For
two functions $F(x)$ and $G(x)$, the coordinate expression for the Moyal
bracket $[F,G]$ is
\begin{equation}
\lbrack F,G]=[x^{a},G]\frac{\partial F}{\partial x^{a}}.
\end{equation}
More details about the origin of the $*$ product and other important
properties are available in literature [9-13]. Next, we will study the
Feynman's proof of Maxwell equations using the NC framework, a fact which
consist also in using the Moyal bracket instead of the Poisson bracket. In
what follows we will present two kind of NC framework associated to two
distinguished scenarios to conceive the proof of the Maxwell equations in a
NC space. These two scenarios offering two possibilities to make the space
NC permit among others to debate the novelty extracted from this extension
relatively to each case.

\subsection{Noncommutativity: First kind}

One way to make the space NC is to consider the following commutation
relations
\begin{equation}
\begin{array}{lcr}
\lbrack x_{j},x_{k}]=i\theta _{jk}, &  &
\end{array}
\end{equation}
\begin{equation}
\begin{array}{lcr}
m[x_{j},\dot{x_{k}}]=i\hbar \delta _{jk}, &  &
\end{array}
\end{equation}
where $[,]$ stands for the Moyal bracket and where (38) is simply
a NC extension of (8). We assume in this first kind of
noncommutativity that the r.h.s of (39) is not affected by the
deformation parameter. Differentiating this equation with respect
to time and using (7) we find the same equation as in the ordinary
product (9), since the NC parameter $\theta _{jk}$ is a constant

\begin{equation}
\lbrack x_{j},F_{k}]+m[\dot{x_{j}},\dot{x_{k}}]=0.
\end{equation}
On the other hand, the bilinearity of the Moyal bracket implies
\begin{equation}
\lbrack [x_{i},F_{j}],x_{k}]+m[[\dot{x_{i}},\dot{x_{j}}],x_{k}]=0.
\end{equation}
Computing the second term of this equation, using the Jacobi identity of $%
\dot{x_{i}}$, $\dot{x_{j}}$ and $x_{k}$ as well as the Moyal bracket of $%
x_{j}$ and $\dot{x_{k}}$ (39), we find the following constraint
\begin{equation}
\lbrack [\dot{x_{i}},\dot{x_{j}}],x_{k}]=0,
\end{equation}
or by virtue of (40)
\begin{equation}
\lbrack [x_{i},F_{j}],x_{k}]=0.
\end{equation}
Compared to the standard computations, the present case shows a new property
namely the quantity $[x_{i},F_{j}]$ is coordinate space independent, and
hence the field $B$ defined by
\begin{equation}
B_{l}=-i\frac{m^{2}}{2{\hbar }}\epsilon _{jkl}[\dot{x_{j}},\dot{x_{k}}]
\end{equation}
is also independent of $x_{i}$. Consequently, the corresponding equations
for $H$ read as
\begin{equation}
divB=\frac{\partial B_{l}}{\partial x_{l}}=0,
\end{equation}
and
\begin{equation}
rotB=\nabla \times B=0.
\end{equation}
Moreover, using (39) and (44), the field E defined by the Lorentz force
equation (10), satisfies then
\begin{equation}
\lbrack x_{m},E_{j}]=0.
\end{equation}
The above equation shows that the field $E$ is also space independent which,
in turn, gives the following equations
\begin{equation}
divE=0,
\end{equation}
and
\begin{equation}
rotE=0.
\end{equation}

Few remarks are in order:

1. The fact to introduce a parameter of noncommutativity to the manner of
(38), induces necessarily the static Maxwell equations which means also the
absence of the charge and current densities $\rho $ and $j$ . We can advance
at this level that the noncommutativity of the first kind is equivalent to
cancel the charge and current densities for the Maxwell equations.

2. It's important to look for the meaning of the commutative limit $\theta
=0.$ In fact, once the previous limit is performed, the behavior of the
Lorentz force $F$ as well as of the field $B$ change completely as they
depend on the behavior of the space coordinates $x_{m}$. Setting $\theta =0$
one discover the Poison bracket $\{x_{j},x_{k}\}=0$ which, by virtue of the
standard computations, means the restoration of the densities $\rho $ and $j$%
.

\subsection{Noncommutativity: Second Kind}

As its shown through the previous calculations, the relation (9) constitutes
a crucial step in the Feynman's proof. Any changes at the level of this
relation will necessarily lead important modifications and all the standards
results are then suspected to change. Here we propose to consider the NC
space (38) and modify the equation (39) while supposing that velocity is a
quantity that depends on spatial coordinates. We suppose the following NC
expressions

\begin{equation}
\begin{array}{lcr}
\lbrack x_{j},x_{k}]=i\theta _{jk}, &  &
\end{array}
\end{equation}
\begin{equation}
\begin{array}{lcr}
m[x_{j},\dot{x_{k}}]=i\hbar \delta _{jk}+im\theta _{jk}f, &  &
\end{array}
\end{equation}
where $f$ is a function which can depend on $x$ and $t$ and it's given by
\begin{equation}
f=\left( \frac{\partial \dot{x_{l}}}{\partial \eta _{l}}(x+\eta )\right)
_{\eta =0.}
\end{equation}
Note that the equation (51) is established by using the definition of the
star product and assuming that
\begin{equation}
\frac{\partial \dot{x_{k}}}{\partial \eta _{b}}=\delta _{kb}\frac{\partial
\dot{x_{l}}}{\partial \eta _{l}}.
\end{equation}
Note also that the velocity should not be proportional to the position due
to the presence of the term $i\hbar \delta _{jk}$ in (51).\newline
However, following step by step the Feynman's analysis [15], one shows that
the derivation of (51) with respect to time $t$ drives naturally to the
following expression
\begin{equation}
m[\dot{x_{j}},\dot{x_{k}}]+m[x_{j},\frac{d\dot{x_{k}}}{dt}]=im\theta _{jk}%
\frac{df}{dt},
\end{equation}
or equivalently
\begin{equation}
m[\dot{x_{j}},\dot{x_{k}}]+[x_{j},F_{k}]=im\theta _{jk}^{{}}\frac{df}{dt}.
\end{equation}
Since the Moyal bracket is also bilinear we can write
\begin{equation}
\lbrack x_{l},[x_{j},F_{k}]]=-m[x_{l},[\dot{x_{j}},\dot{x_{k}}]]+im\theta
_{jk}[x_{l},\frac{df}{dt}].
\end{equation}
Furthermore, using the Jacobi identity of $x_{l}$, $\dot{x_{j}}$ and $\dot{%
x_{k}}$, the first term in the right side of equation (56) gives
\begin{equation}
\lbrack x_{l},[\dot{x_{j}},\dot{x_{k}}]]=i[(\theta _{lk}\dot{x_{j}}-\theta
_{lj}\dot{x_{k}}),f]
\end{equation}
and we can write
\begin{equation}
\lbrack x_{l},[x_{j},F_{k}]]=-im[(\theta _{lk}\dot{x_{j}}-\theta _{lj}\dot{%
x_{k}}),f]+im\theta _{jk}[x_{l},\frac{df}{dt}].
\end{equation}
Note that, in spite of the fact that (55) extends the standard relation (15)
it preserves the antisymmetry property of $x_{j}$ and $F_{k}$, because of
the antisymmetry of the NC parameter $\theta $, namely
\begin{equation}
\lbrack x_{j},F_{k}]=-[x_{k},F_{j}],
\end{equation}
and therefore the field $B$ can also be defined as
\begin{equation}
\lbrack x_{j},F_{k}]=-(i\hbar /m)\epsilon _{jkl}B_{l}.
\end{equation}
Equations (58) and (60) give the following Moyal bracket
\begin{equation}
\lbrack x_{l},B_{s}]=\frac{m^{2}}{2{\hbar }}\epsilon _{jks}\left( [(\theta
_{lk}\dot{x_{j}}-\theta _{lj}\dot{x_{k}}),f]-\theta _{jk}[x_{l},\frac{df}{dt}%
]\right) ,
\end{equation}
which vanishes for $\theta =0$, giving rise then to the standard Poisson
bracket (20).\newline
The field $B$ can be written using (55)and (60) as follows
\begin{equation}
B_{s}=-i\frac{m^{2}}{2{\hbar }}\epsilon _{jks}[\dot{x_{j}},\dot{x_{k}}]-%
\frac{m^{2}}{2{\hbar }}\epsilon _{jks}\theta _{jk}\frac{df}{dt}.
\end{equation}
Note that the second term in the right hand side of (62) didn't appear in
standard calculations (22). On the other hand, using (60) as well as the
expression of the Lorentz force (10) we can write for the electric field $E$%
\begin{equation}
\lbrack x_{j},E_{k}]=-\epsilon _{kmn}\dot{x_{m}}[x_{j},B_{n}]-i\epsilon
_{kmn}\theta _{jm}fB_{n},
\end{equation}
To explicit much more this expression one need only to substitute the
bracket $[x_{j},B_{n}]$ and $B_{n}$ by their explicit formulas (61-62). Now,
in order to obtain the NC analogous of the first Maxwell equation $divB=0$,
one should compute, as previously, the Moyal bracket between the velocity
and the field $B$
\begin{equation}
\lbrack \dot{x_{s}},B_{s}]=-i\frac{m^{2}}{2{\hbar }}\epsilon _{jks}[\dot{%
x_{s}},[\dot{x_{j}},\dot{x_{k}}]]-\frac{m^{2}}{2{\hbar }}\epsilon
_{jks}\theta _{jk}[\dot{x_{s}},\frac{df}{dt}],
\end{equation}
or simply
\begin{equation}
\lbrack B_{s},\dot{x_{s}}]=\frac{m^{2}}{2{\hbar }}\epsilon _{jks}\theta
_{jk}[\dot{x_{s}},\frac{df}{dt}],
\end{equation}
since the first term of (64) vanishes using the analogous of the Jacobi
identity(23).\newline
Afterwards, using (37), this equation becomes
\begin{equation}
(i\hbar \delta _{as}+im\theta _{as}f)\frac{\partial B_{s}}{\partial x_{a}}=%
\frac{m^{3}}{2{\hbar }}\epsilon _{jks}\theta _{jk}[\dot{x_{s}},\frac{df}{dt}%
],
\end{equation}
or equivalently
\begin{equation}
\frac{\partial B_{s}}{\partial x_{s}}=-i\frac{m^{3}}{2{\hbar }^{2}}\epsilon
_{jks}\theta _{jk}[\dot{x_{s}},\frac{df}{dt}]-\frac{m}{{\hbar }}\theta _{as}f%
\frac{\partial B_{s}}{\partial x_{a}}.
\end{equation}
Using once again (61) and the following identity
\begin{equation}
\lbrack B_{s},x_{s}]=[x_{a},x_{s}]\frac{\partial B_{s}}{\partial x_{a}}%
=i\theta _{as}\frac{\partial B_{s}}{\partial x_{a}},
\end{equation}
the first NC Maxwell equation corresponding to (51) reads finally as
\begin{eqnarray}
div_{\theta }B &=&\frac{\partial B_{s}}{\partial x_{s}}=-i\frac{m^{3}}{2{%
\hbar }^{2}}\epsilon _{jks}\theta _{jk}[(\dot{x_{s}}+2x_{s}),\frac{df}{dt}]
\nonumber \\
&+&i\frac{m^{3}}{{\hbar }^{2}}\epsilon _{jks}[(\theta _{ks}\dot{x_{j}}%
-\theta _{sj}\dot{x_{k}}),f].
\end{eqnarray}
This equation can be simply rewritten as
\begin{equation}
div_{\theta }B=\eta _{\theta },
\end{equation}
where we have introduced the notation $div_{\theta }B\equiv \frac{\partial
B_{s}}{\partial x_{s}}$ for the first NC Maxwell equation to distinguish it
from the standard case. A remarkable fact is that the r.h.s. of (69) namely $%
\eta _{\theta },$ is completely dependent of the NC parameter $\theta $,
setting $\theta =0$ we obtain exactly the ordinary Maxwell equation (24).
Here, one could anticipate and give a significance to this new immersing
term $\eta _{\theta }$ as being a density of magnetic charges in analogy
with the density of electric charge.\newline
Next, to obtain the second NC Maxwell equation, we derive with respect to
time the field $B_{s}$ (62)

\begin{equation}
\frac{\partial B_{s}}{\partial t}+\dot{x_{m}}\frac{\partial B_{s}}{\partial
x_{m}}=-i\frac{m^{2}}{{\hbar }}\epsilon _{jks}[\frac{d\dot{x_{j}}}{dt},\dot{%
x_{k}}]-\frac{m^{2}}{{2\hbar }}\epsilon _{jks}\theta _{jk}\frac{d^{2}f}{%
dt^{2}},
\end{equation}
this is because the magnetic field $B$ is $(x,t)$-coordinates dependent,
since the velocity is also considered as depending on the space coordinate.
Furthermore, using the Lorentz force (10), one have
\begin{eqnarray}
\frac{\partial B_{s}}{\partial t}+\dot{x_{m}}\frac{\partial B_{s}}{\partial
x_{m}} &=&-i\frac{m}{{\hbar }}\epsilon _{jks}[E_{j}+\epsilon _{jmn}\dot{x_{m}%
}B_{n},\dot{x_{k}}]  \nonumber \\
&&-\frac{m^{2}}{{2\hbar }}\epsilon _{jks}\theta _{jk}\frac{d^{2}f}{dt^{2}}
\nonumber \\
&=&-i\frac{m}{{\hbar }}\left( \epsilon _{jks}[E_{j},\dot{x_{k}}]+[\dot{x_{k}}%
B_{s},\dot{x_{k}}]-[\dot{x_{s}}B_{k},\dot{x_{k}}]\right) \\
&&-\frac{m^{2}}{{2\hbar }}\epsilon _{jks}\theta _{jk}\frac{d^{2}f}{dt^{2}},
\nonumber
\end{eqnarray}
Explicitly we find the following expression for the second NC Maxwell
equation
\begin{eqnarray}
\frac{\partial B_{s}}{\partial t}+\epsilon _{kjs}\frac{\partial E_{j}}{%
\partial x_{k}} &=&-i\frac{m}{{\hbar }}\epsilon _{jks}f[E_{j},x_{k}]-i\frac{m%
}{{\hbar }}\dot{x_{k}}f[B_{s},x_{k}]  \nonumber \\
&+&\dot{x_{s}}[\dot{x_{k}},B_{k}]-\frac{m^{2}}{{2\hbar }}\epsilon
_{mnk}\theta _{mn}[\dot{x_{k}},\dot{x_{s}}]\frac{df}{dt},
\end{eqnarray}
leading then, after some algebraic manipulations, to the following compact
formula
\begin{equation}
\frac{\partial B_{s}}{\partial t}+\epsilon _{kjs}\frac{\partial E_{j}}{%
\partial x_{k}}=A_{1}\frac{d^{2}f}{dt^{2}}+A_{2}\frac{df}{dt}+A_{3}
\end{equation}
where the r.h.s. term of (74) is a non linear second order differential
equation in the arbitrary function $f$ whose coefficients are explicitly
given by
\begin{eqnarray}
A_{1} &=&-\frac{m^{2}}{{2\hbar }}\epsilon _{jks}\theta _{jk}  \nonumber \\
A_{2} &=&\frac{m^{3}}{{2\hbar }^{2}}\theta _{jl}\left\{ \theta _{ks}\epsilon
_{jls}f^{2}-i\epsilon _{jlk}[\dot{x_{s}},\dot{x_{k}}]\right\} \\
A_{3} &=&\frac{m^{3}}{{2\hbar }^{2}}\left\{ \epsilon _{jlk}[\dot{x_{s}},\dot{%
x_{k}}]+i\theta _{ks}\epsilon _{jls}f^{2}\right\} [\dot{x_{j}},\dot{x_{l}}]-%
\dot{x_{s}}\eta _{\theta }.  \nonumber
\end{eqnarray}
where $\eta _{\theta }=div_{\theta }B$ (69-70). As we can easily check, all
the local coefficients functions $A_{1}$, $A_{2}$ and $A_{3}$ are $\theta $%
-depending. Thus, the standard limit $\theta =0$ is natural as it leads to
the standard Feynman's proof computations. Our last forthcoming section is
devoted to a conclusion with a discussion about the derived results.

\section{Concluding Remarks.}

\smallskip

Let us summarize what has been the scope of this work. The importance of the
so called Feynman's proof of the Maxwell equations was essentially revealed
by the Dyson's work [15]. This paper resuscitated a former idea of Feynman
who made a proof of the Maxwell equations assuming only the Newton's law of
motion and the commutation relations between position and velocity. This
proof that Feynman refused to publish, believing that it was a simple joke
[21], was appreciated and taken with a great seriousness by several
scientists [16-19].\newline

However, one of the things that caused some discussions around the Feynman's
proof is the fact that the derivation mixes classical and quantum concepts
and the small confusion that seems to appear when we see the relativistic
Maxwell equations derived from the classical Newton's law. The point is that
the consideration of non relativistic equations and the relations of
commutation between position and velocity are only a process well arranged
to get the Maxwell equations. As it is signaled in [18], one may wonder then
how truly relativistic Maxwell equations are derived from Newton's classical
assumptions?. The confusion could be shaped if we remark that the Feynman's
proof concerns only half of Maxwell equations, namely $divB=0$ and $\frac{%
\partial B}{\partial t}+\nabla \times E=0,$ describing the Bianchi set of
equations. It doesn't anymore pose problem since this couple of equations is
compatible with the nonrelativistic Galilean invariance.\newline

Following the Feynman's proof of the Maxwell equations, assuming only the
Newton's law of motion and the commutation relation between position and
velocity, we try in this paper to study this proof using the NC geometry
framework. To accomplish this task, we consider two kinds of NC formulations
going along the same way as Feynman's approach. This allows us to discover,
in a first formulation, the static Maxwell equations. Afterwards, motivated
by the hope to find a new theory using NC framework, we assume that the
velocity is also space dependent and write the modified NC relation (51).
The results extracted from the second formulation are more significant as
they are associated to a non trivial $\theta $-extension of the Bianchi-set
of Maxwell equations namely $div_{\theta }B=\eta _{\theta }$ and $\frac{%
\partial B_{s}}{\partial t}+\epsilon _{kjs}\frac{\partial E_{j}}{\partial
x_{k}}=A_{1}\frac{d^{2}f}{dt^{2}}+A_{2}\frac{df}{dt}+A_{3},$ where $A_{1}$, $%
A_{2}$ and $A_{3}$ are local coefficient functions depending on the NC
parameter $\theta $. Our objectives in reconsidering the Feynman's proof
are, on one hand, to put it in relief and, on the other hand, to show its
importance in the NC framework.\newline

The novelty of this proof in the NC space is revealed notably at the level
of the corrections brought to the previous Maxwell equations. These
corrections correspond essentially to the possibility of existence of
sources of magnetic charges that we can associate to the magnetic monopole
since $div_{\theta }B=\eta _{\theta }$. Note that these extra terms $\eta
_{\theta }$ are absent in the ordinary case associated to $\theta =0.$ These
results may give new insights into the study of the electromagnetic duality
and its various physical and mathematical aspects.\newline

{\bf Acknowledgments}

\smallskip

We would like to thank the Abdus Salam International Centre for Theoretical
Physics, Trieste-Italy, and the considerable help of the High Energy Section
where this work was done. We acknowledge Dr. M. Hssaini for fruitful
discussions. This work was done within the framework of the Associateship
scheme of the Abdus Salam ICTP (Summer 2003).

\newpage {\bf References}

\begin{enumerate}
\item[\lbrack 1]  ] A.Connes, Academic Press (1994).

\item[\lbrack 2]  ] N.Seiberg, E.Witten, ''String Theory and Noncommutative
Geometry'', JHEP 9909, 032 (1999), hep-th/9908142.

\item[\lbrack 3]  ] A.Connes, M.R.Douglas, A.Shwartz, ''Noncommutative
Geometry and Matrix Theory, Compactification on Tori'', JHEP 9802, 003
(1998), hep-th/9711162.

\item[\lbrack 4]  ] E. Witten, ''NC Tachyons And String Field Theory'',
hep-th/0006071.

\item[\lbrack 5]  ] M.M.Seikh-Jabbari, ''SuperYang-Mills Theory on NC Torus
from Open Strings Interactions'', Phys. Lett. B 450 (1999) 119,
hep-th/9810179; Ihab. F. Riad, M.M. Sheikh-Jabbari, ''NC QED and Anomalous
Dipole Moments'', JHEP 0008 (2000) 045, hep-th/0008132; A. Micu, M.M.
Sheikh-Jabbari, ``Noncommutative $\Phi ^{4}$ Theory at Two Loops'', JHEP
0101 (2001) 025, hep-th/0008057

\item[\lbrack 6]  ] L.Alvarez-Gaume, J.F.Barbon, ''Non-linear Vacuum
Phenomena in NC QED, hep-th/0006209.\newline
R. J. Szabo, ``Quantum Field Theory on Noncommutative Spaces'', Phys.Rept.
378 (2003) 207-299.

\item[\lbrack 7]  ] N.Ishibashi, ''A Relation between Commutative and NC
Descriptions of D-branes'', hep-th/9909176, \newline
F.Ardalan, H.Arfaei, M.M.Seikh-Jabbari, ''NC Geometry from Strings and
Branes'' JHEP 9902 (1999) 016,\newline
Pei-Ming Ho, Yong-Shi Wu, ``Noncommutative Geometry and D-Branes'',
Phys.Lett. B398 (1997) 52-60,

\item[\lbrack 8]  ] H.S. Snyder, ``Quantized Space-Time'', Phys.Rev. 71
(1947) 38,

\item[\lbrack 9]  ] M. Kontsevitch, ``Deformation quantization of Poisson
manifolds I'', q-alg/9709040.

\item[\lbrack 10]  ] D.B. Fairlie, ``Moyal Brackets, Star Products and the
Generalized Wigner Function'', hep-th/9806198\newline
D.B. Fairlie, ``Moyal Brackets in M-Theory'', Mod.Phys.Lett. A13 (1998)
263-274, hep-th/9707190,\newline
C. Zachos, ``A Survey of Star Product Geometry'', hep-th/0008010;

C. Zachos, ``Geometrical Evaluation of Star Products'', J.Math.Phys. 41
(2000) 5129-5134,\newline
C. Zachos, T. Curtright, ``Phase-space Quantization of Field Theory'',
Prog.Theor. Phys. Suppl. 135 (1999) 244-258.

\item[\lbrack 11]  ] A. Das and Z. Popowicz, ``Properties of Moyal-Lax
Representation'', Phys.Lett. B510 (2001) 264-270, hep-th/0103063. \newline
A. Das and Z. Popowicz, ``Supersymmetric Moyal-Lax Representations'', J.
Phys. A 34(2001) 6105 and hep-th/0104191.

\item[\lbrack 12]  ] Ming-Hsien Tu, Phys.Lett. B508 (2001) 173-183,\newline
Ming-Hsien Tu, Niann-Chern Lee, Yu-Tung Chen, J.Phys. A35 (2002) 4375.

\item[{\lbrack 13]}]  A. Boulahoual and M.B. Sedra, ``Non Standard Extended
Noncommutativity of Coordinates'', hep-th/0104086;

A. Boulahoual and M.B. Sedra, ``The Das-Popowicz Moyal Momentum Algebra'',
hep-th/0207242;

A. Boulahoual and M.B. Sedra, ``The Moyal Momentum algebra applied to $%
\theta $-deformed 2d conformal models and KdV-hierarchies'', hep-th/0208200.

\item[{\lbrack 14]}]  S. Ghosh, ``Effective Field Theory for Noncommutative
Spacetime'': A Toy Model'', hep-th/0307227.

S. Ghosh, ``Batalin-Tyutin Quantisation of the Spinning Particle Model'',
hep-th/0103258, J.Math.Phys. 42 (2001) 5202-5211.

S. Ghosh, ``Covariantly Quantized Spinning Particle and its Possible
Connection to Non-Commutative Space-Time'', Phys.Rev. D66 (2002) 045031,
hep-th/0203251

\item[\lbrack 15]  ] F.J. Dyson, ''Feynman's proof of the Maxwell
equations'', (1990). American Journal of Physics, 58, 209.

\item[\lbrack 16]  ] P.Bracken, ''Poisson Brackets and the Feynman
Problem'', International Journal of Theoretical Physics, Vol. 35, No. 10,
1996.

\item[\lbrack 17]  ] P.Bracken, ''Relativistic equations of motion from
Poisson brackets'', International Journal of Theoretical Physics, Vol. 37,
No. 5, 1998.

\item[\lbrack 18]  ] Z. K. Silagadze, ``Feynman's derivation of Maxwell
equations and extra dimensions'' Annales de la fondation Louis de Broglie,
Volume 27 No2, 2002, hep-ph/0106235.

\item[\lbrack 19]  ] A. Berard, H. Mohrbach and P. Gosselin,
``Lorentz-Cocariant hamiltinain formalism'', Physics/0004005,
Int.J.Theor.Phys. 39 (2000) 1055-1068.

A. Berard, y. Grandati, H. Mohrbach, ``Dirac monopole with Feynman
brackets'', Phys.Lett. A254 (1999) 133-136, physics/0004008.

\item[{\lbrack 20]}]  R. Penrose, Journal of mathematical Physics, 10
(1969)38.

\item[{\lbrack 21]}]  F.J. Dyson, ''Feynman at Cornell'', Phys. Today 42
(2), 32-38 (1989).
\end{enumerate}

\end{document}